\documentclass[twocolumn]{article} 
\usepackage{graphicx} 

\usepackage{amsmath,tikz}
\usetikzlibrary{arrows,automata,positioning}

\long\def\comment#1{}

\makeatletter
\def\@normalsize{\@setsize\normalsize{10pt}\xpt\@xpt
\abovedisplayskip 10pt plus2pt minus5pt\belowdisplayskip
\abovedisplayskip \abovedisplayshortskip \z@
plus3pt\belowdisplayshortskip 6pt plus3pt
minus3pt\let\@listi\@listI}

\def\subsize{\@setsize\subsize{12pt}\xipt\@xipt}
\def\section{\@startsection {section}{1}{\z@}{1.0ex plus
1ex minus .2ex}{.2ex plus .2ex}{\large\bf}}
\def\subsection{\@startsection
   {subsection}{2}{\z@}{.2ex plus 1ex} {.2ex plus .2ex}{\subsize\bf}}
\makeatother

\begin{document}

\date{}

\title{\huge \bf {Modeling Super-spreading Events for Infectious Diseases: Case Study SARS}}

\author{Thembinkosi Mkhatshwa
\thanks{Email: tpsmjc@yahoo.com Tel: 865-805-4292.  Both authors were supported by the MU-ADVANCE Faculty Fellowship Award, Marshall University, Huntington, WV.
 }
\and
Anna Mummert
 \thanks{Corresponding author.
 Marshall University, Mathematics Department, Huntington, WV USA 25755
 Tel/Fax: 304-696-3041/4646 Email: mummerta@marshall.edu. Manuscript submitted \today
 }
}

\maketitle
\thispagestyle{empty}


{\hspace{1pc} {\it{\small Abstract}}{\bf{\small---Super-spreading events for infectious diseases occur when some infected individuals infect more than the average number of secondary cases.  Several super-spreading individuals have been identified for the 2003 outbreak of severe acute respiratory syndrome (SARS).  We develop a model for super-spreading events of infectious diseases, which is based on the outbreak of SARS.  Using this model we describe two methods for estimating the parameters of the model, which we demonstrate with the small-scale SARS outbreak at the Amoy Gardens, Hong Kong, and the large-scale outbreak in the entire Hong Kong Special Administrative Region.  One method is based on parameters calculated for the classical susceptible - infected - removed ($SIR$) disease model.  The second is based on parameter estimates found in the literature.  Using the parameters calculated for the $SIR$ model, our model predicts an outcome similar to that for the $SIR$ model.  On the other hand, using parameter estimates from SARS literature our model predicts a much more serious epidemic.

\em Keywords: modeling, infectious diseases, super-spreading events,  severe acute respiratory syndrome }}
 }


\section{Introduction}
Severe acute respiratory syndrome (SARS) is a highly contagious respiratory disease which is caused by the SARS Coronavirus. It is a serious form of pneumonia, resulting in acute respiratory distress and sometimes death. SARS emerged in China late 2002 and quickly spread to 32 countries causing more than 774 deaths and 8098 infections worldwide. One of the intriguing characteristics of the SARS epidemic was the occurrence of super-spreading events. Super-spreading events for a specific infectious disease occur when certain infected individuals, called super-spreaders, infect more than the average number of secondary cases.

According to the U.\ S.\ Centers for Disease Control and Prevention, a person is a super-spreader if they cause more than 10 secondary infections.  Such super-spreading individuals have been identified in the SARS outbreak and they are thought to have caused most of the secondary infections.  For example, in Singapore, about 80\% of infections have been attributed to only 5 super-spreading individuals~(\cite{Leo2003}).  One extreme super-spreading individual in Hong Kong caused more than 100 secondary infections~(\cite{Riley2003}).  To contrast, in Singapore, most individuals caused 0 secondary infections~(\cite{Masuda2004}).

One differential equation model for super-spreading individuals was proposed before the recent SARS outbreak, by Kemper~(\cite{Kemper1980}).  He presents a modified  susceptible - infected - removed ($SIR$) disease model to capture the effect of the super-spreading individuals in which the infected individuals are split into two different infected classes with different transmission rates.  The super-spreaders have a higher transmission rate, meaning more of their contacts with susceptible individuals result in a new infection than the regular infected individuals.  This was the first model designed specifically to address the effect of super-spreading individuals on the course of infectious disease epidemics.

Due to the extreme influence of super-spreading individuals during the SARS outbreak, many mathematical epidemiologists developed models for the spread of SARS, which included the occurrence of the super-spreading events\footnote{Many models were developed to try and mathematically capture the SARS outbreak.  Interested readers are directed to Bauch, et al.~(\cite{Bauch2005}) for an overview of SARS models.}.  It is interesting to note how each incorporated the super-spreading events into their model.  First are models where the super-spreading individuals have a higher transmission rate than the regular infected individuals, as in Kemper~(\cite{Kemper1980}).  For example, Masuda, et\ al.~(\cite{Masuda2004}) developed a contact network model which had different transmission rates for regular infected individuals and for super-spreading individuals.  Second are models in which parameters are taken from probability distributions, making the super-spreading individuals naturally appear as the right-hand tail of the distributions.  Lloyd-Smith, et\ al.~(\cite{Lloyd-Smith2005}) developed a stochastic compartment model and Fang, et\ al.~(\cite{Fang2004}) developed a spacial lattice combined with a deterministic compartment model wherein the individual reproduction numbers are drawn from a continuous probability distribution.

The goal of this paper is to capture the effect of the super-spreading individuals using a modification of the classical $SIR$ disease model.  The modification is inspired by Li, et\ al.~(\cite{Li2004}) who determined  that for SARS ``the daily infection rate did not correlate with the daily total number of symptomatic cases but with the daily number of symptomatic cases who were not admitted to a hospital within 4 days of onset of symptoms."  This means that the number of infected individuals is closely associated with the number of individuals who remain outside of isolation longer than most other infected individuals.  These individuals remaining outside of isolation longer than normal are related to the disease severity.  They have more contacts with susceptible individuals and thus more chances to spread the disease, in other words, they are the super-spreaders.  In our model, we split the infected individuals into two classes, with two different ``removal" rates; these two rates determine how long an infected individual remains outside of isolation.

In this paper, we modify the classical $SIR$ disease model to capture the effect of a super-spreading event, using the idea that super-spreading individuals stay out of isolation longer than individuals who are not super-spreading.
A description of the $SIPR$ model, the model equations, and some basic model properties are given in Section~\ref{SIPRmodel}.  We demonstrate the model using a small-scale and a large-scale outbreak of SARS in Section~\ref{parameterestimation}.
Our discussion follows in Section~\ref{discussion}.

\subsection{The $SIR$ model}
We conclude the introduction with a brief review of the classical $SIR$ disease model.  A more detailed review is given in Murray~(\cite{Murray}), and in Ching, et al.~(\cite{Ching2003}) and the references given there.

The classical $SIR$ disease model splits a fixed-size population, $N$, into three distinct classes: the susceptible individuals, $S$, those do not have the disease and can become infected; the infected individuals, $I$, those who have the disease and can infect susceptible individuals; and the removed individuals, $R$, those who have recovered, die, or moved into isolation.  Individuals in the removed class gain permanent immunity and remain in the $R$ class forever.  Schematically, the $SIR$ model is
\[ S \longrightarrow I \longrightarrow R \]
Individuals in the population are assumed to homogeneously mix.  Contacts between the susceptible and infected individuals result in a new infected individual at a rate proportional to the to the number of susceptible and infected individuals.  Infected individuals are removed to class $R$ at a rate proportional to the number of infected individuals.  The system of nonlinear ordinary differential equations describing the $SIR$ model is
\begin{eqnarray*}
\label{SIRequations}
\frac{dS}{dt}& = &-\beta SI\notag\\
\frac{dI}{dt}& = &\beta SI-\nu I\\
\frac{dR}{dt}& = &\nu I\notag
\end{eqnarray*}
where $\beta$ is the transmission rate and $\nu$ is the removal\footnote{The $R$ class is often called the recovered class containing those who have recovered from the disease.  In this situation, all individuals recover from the disease; no individuals die or move into isolation. The parameter $\nu$ is correspondingly called the recovery rate.} rate.

\section{The $SIPR$ model}
\label{SIPRmodel}
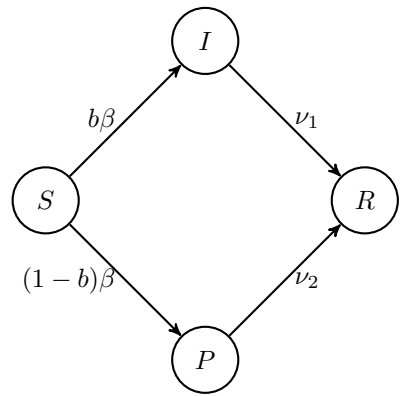
\begin{figure}
\begin{center}
\begin{tikzpicture}[->,>=stealth',node distance=3cm,
                 thick]
\tikzstyle{every state}=[draw=black,thick]

\node[state]         (A)                    {$S$};
\node[state]         (B) [above right of=A] {$I$};
\node[state]         (D) [below right of=A] {$P$};
\node[state]         (C) [below right of=B] {$R$};

\path (A) edge              node [left] {$b\beta$}      (B);
\path (A) edge              node [left]  {$(1-b)\beta$} (D);
\path (B) edge              node [right] {$\nu_1$}(C);
\path (D) edge              node [right] {$\nu_2$}(C);
\end{tikzpicture}
\end{center}
\caption{$SIPR$ model schematic, where $S$ are the susceptible individuals, $I$ are the regular infected individuals, $P$ are the super-spreaders, $R$ are the removed individuals, $\beta$ is the transmission rate, $b$ is the probability that a new infection will be a regular infected person, $1-b$ is the probability that a new infection will be a super-spreading individual, $\nu_1$ is the removal rate for a regular infected individual, and $\nu_2$ is the removal rate for a super-spreading individual.}
\label{SIPRSchematic}
\end{figure}

We describe the $SIPR$ model, a modification of the classical $SIR$ model, which captures the effect of super-spreading individuals. Schematically, the model is given in Figure~\ref{SIPRSchematic}.
While the basic $SIR$ model has one class of infected individuals, the $SIPR$ model has a second class of infected individuals, the super-spreaders, denoted by the variable $P$.
In the $SIPR$ model we divide the population, of fixed size $N$, into four groups, namely the susceptible individuals, $S$,
 the regular infected individuals, $I$,  the super-spreaders, $P$, and the removed individuals, $R$.

 A susceptible individual can become infected through contact with either a regular infected individual or a super-spreading individual. Then with probability $b$ the new infected individuals become regular infected individuals (move to class $I$) and with probability $1-b$ the new infected individuals become super-spreading individuals (move to class $P$).  We assume most new infections are regular infected individuals, that is, $b>1-b$.
We assume that the two infected classes, $I$ and $P$, have the same transmission rate, $\beta$.

To capture the effect of the super-spreading individuals, we use the idea that individuals who are super-spreaders stay out of isolation longer than the regular infected individuals.  The regular infected individuals stay out of isolation (the $R$ class) for $1/\nu_1$ days and the super-spreading individuals stay out of isolation for $1/\nu_2$ days, with $1/\nu_1 < 1/ \nu_2$.  Therefore, the model has two distinct removal rates namely, $\displaystyle{\nu_1}$ and $\displaystyle{\nu_2}$, corresponding to regularly infected individuals, $I$, and super-spreading events, $P$, respectively.
When a person is removed to the $R$ class there is no possibility of becoming susceptible again, but rather they recover and gain permanent immunity, or die; in either case, they remain in the removed compartment forever.

Based on the previous descriptions and assumptions we formulate a system of four ordinary differential equations for the $SIPR$ model.
\begin{eqnarray}
\label{SIPRequations}
\frac{dS}{dt}& = &-\beta (I+P)S\notag\\
\frac{dI}{dt}& = &b\beta (I+P)S-\nu_1I\\
\frac{dP}{dt}& = &(1-b)\beta (I+P)S-\nu_2P\notag\\
\frac{dR}{dt}& = &\nu_1I+\nu_2P\notag
\end{eqnarray}
subject to the following initial conditions
\[S(0) = N - I_0 -P_0 = S_0,\] \[ I(0) = I_0,\quad  P(0) = P_0, \quad \text{and} \quad R(0) = 0.\]
The $SIPR$ model given by Equations~(\ref{SIPRequations}) has the following properties.
\begin{itemize}
\item The $SIPR$ model has a unique global solution.
\item The components of the solution,~$S(t)$,~$I(t),P(t)$, and $R(t)$, of the $SIPR$ model are non-negative and bounded by $N$ for all time, $t \geq 0$.
\item The $SIPR$ model has equilibrium points $(N,0,0,0)$, $(S^*,0,0,R^*)$, for any $0<S^*<S_0$, and $(0,0,0,N)$.
\item The individual reproduction number is
\[R_0= \frac{b \beta S_0}{\nu_1} + \frac{(1-b) \beta S_0}{\nu_2}.\]
\end{itemize}
The properties of the $SIPR$ are, in general, the natural modification of the corresponding properties of the classical $SIR$ model.  It is interesting to note that the $R_0$ for the $SIPR$ model is the $R_0$ for the corresponding $SIR$ model for each of the two infected classes, $I$ and $P$, multiplied by the probability of a new infected individual becoming an $I$ or a $P$, respectively.

The $SIPR$ model can be used to analyze any infectious disease where super-spreading events have been identified. As a particular example, super-spreading events have been identified in one outbreak of measles as described in Paunio, et al.~(\cite{Paunio1998}).

\section{The SARS Epidemic}
\label{parameterestimation}

We use the $SIPR$ model to study the spread of SARS on a small scale, in the Amoy Gardens apartment complex in Hong Kong, and on a large scale, in the entire Hong Kong Special Administrative Region.  In both cases, we fit the model to the data using two parameter estimation methods.  We begin with a description of the estimation procedures.

The general parameter estimation procedure was as follows.  The $SIPR$ system of equations~(\ref{SIPRequations}) was solved repeatedly with parameter sets taken from the allowable range of possible parameters, and the least squares error between the cumulative number of cases of the solution and the actual data was computed.  The least squares error was minimized.  All computations were done with MATLAB using the function fminsearchbnd.

In parameter estimation Method 1, we used the parameters estimated with a fit of the classical $SIR$ model to the data.  For the data on the entire Hong Kong Special Administrative Region, we were unable to find $SIR$ parameters.  Therefore, we began with fitting the classical $SIR$ model to the data.  A summary of the $SIR$ model parameters is given in Table~\ref{SIRTable}.  We use the transmission rate of the $SIR$ model as the transmission rate of the $SIPR$ model.  The transmission rate, the removal rate, and the initial number of susceptible individuals are used to determine the basic reproduction number $R_0$.  We estimate parameters $b$, $\nu_1$, and $\nu_2$ so that together they have $R_0$ as calculated, and they satisfy $b > 1-b$ and $\nu_1>\nu_2$.

\begin{table*}
\begin{center}
\begin{tabular}{|c|ccc|} \hline
Parameter & Amoy Gardens & Hong Kong & Hong Kong \\
 & & Feb 21 - Mar 21 & Mar 29 - Jun 12 \\\hline
$\beta$ & 1.4850/N&0.2586/N & 0.1351/N\\
$\nu$ &0.9750 &0.0821 & 0.1923\\
$R_0$ &1.5154 &3.1511 & 0.7025\\ \hline
\end{tabular}
\end{center}
\caption{Estimated parameters for the $SIR$ model and the SARS outbreak in the Amoy Gardens, Hong Kong~(from (\cite{Ching2003})), and in the Hong Kong Special Administrative Region for both February 21 - March 21, 2003 and March 29 - June 12, 2003.}
\label{SIRTable}
\end{table*}

In parameter estimation Method 2, bounds for the parameters in the $SIPR$ model were determined from the literature.  Then, using these bounds, we estimate the parameters that give the best fit to the data.  Details and references for these bounds are given in the next paragraph and in Subsections~\ref{fitmodeltodata} and~\ref{SARSHK}.

For estimation Method 2, we make some assumptions regarding the parameters that apply to both the small- and large-scale outbreaks.  In situations with super-spreading individuals, it is assumed that most infected individuals are not super-spreaders, which is the case for the spread of SARS.  In fact, Masuda, et al.~(\cite{Masuda2004}) state that for the SARS outbreak in Singapore 80\% of infected individuals infected no one else. This leads us to set a lower bound on the probability of becoming a regular infected individual of 0.8, and we have $0.8<b<1$.  For SARS, after onset of symptoms regular infected individuals stayed out of isolation between 3 and 5 days~(\cite{Donnelly2003}); we assume the average and set $\nu_1 = 1/4$.
We assume that on average super-spreading individuals moved into isolation 10 days after they became infectious, that is, $\nu_2=1/10$.
Finally, many researcher have shown that the individual reproduction number, $R_0$, for the SARS outbreak is between 1.5 and 4~(\cite{Bauch2005}).  We assume this region for $R_0$, which gives a constraint on the parameters $\beta$, $b$, $\nu_1$, and $\nu_2$.

\subsection{The SARS epidemic - small scale}
\label{fitmodeltodata}

We use the $SIPR$ model to study the spread of SARS at the high rise apartment building Amoy Gardens, Block~E, in Hong Kong.  We assume that the cumulative number of confirmed cases of SARS in the Amoy Gardens, summarized in Table~\ref{AmoyGardensTable}, are all from Block~E, and that Block~E contains a total of 792 individuals, as in Ching, et al.~(\cite{Ching2003}).

For parameter estimation Method 1, we use the parameters determined by Ching, et al.~(\cite{Ching2003}).  The fixed and estimated parameters of the $SIPR$ model are given in Table~\ref{ParameterTableSARS}.  The resulting number of confirmed cases of SARS are presented in Table~\ref{AmoyGardensTable}.

\begin{table*}
\begin{center}
\begin{tabular}{|c|ccccc|} \hline
Day & March 26 & March 27 & March 28 & March 29 & March 30 \\ \hline
Confirmed & 7 & 22 & 56 & 78 & 114  \\
Method 1 & 12.2&26.2 &48.2 &80.3  &123.8  \\
Method 2  &9.1 &19.1 &37.4 &69.7  &122.3 \\ \hline
\end{tabular}
\end{center}
\caption{Cumulative confirmed cases of SARS in Amoy Gardens, Block~E, March 2003 (from Ching, et al.~(\cite{Ching2003})).  Predicted cases based on parameter estimation Methods~1 and~2.}
\label{AmoyGardensTable}
\end{table*}

\begin{table*}
\begin{center}
\begin{tabular}{|cc|cc|cc|} \hline
Parameter &Dimension & Method 1 &  Fixed  or Estimated & Method 2 & Bound or Constraint\\ \hline
$S_0$ &people & 788 & Fixed &788 &  $S_0$=788 \\
$I_0$ &people & 0.4 & Estimated & 0.0
 &  $I_0 + P_0 = 4$\\
$P_0$ &people & 3.6 & Fixed by $I_0$ & 4.0  & $I_0 + P_0 = 4$\\
$\beta$ &days$^{-1} \times$ people$^{-1}$& 1.4850/N & Fixed &  0.8728/N &$0<\beta < 0.01$ \\
$b$ & & 0.9315 &Estimated & 0.9994  &$0.8<b<1$\\
$\nu_1$ &days$^{-1}$& 0.9997 & Estimated & 0.2500& $\nu_1=0.25$\\
$\nu_2$ & days$^{-1}$& 0.7294& Fixed by $b$, $\nu_1$, and $R_0$ & 0.1000 &$\nu_2 =0.1$\\
$R_0$ & & 1.5154  &Fixed& 3.4765  & $1.5 < R_0 < 4$\\ \hline
\end{tabular}
\end{center}
\caption{Estimated parameters for the $SIPR$ model and the SARS outbreak in the Amoy Gardens, Hong Kong, March 2003, using Methods 1 and 2.  Fixed parameters were found in Ching et al.~(\cite{Ching2003}).}
\label{ParameterTableSARS}
\end{table*}

For parameter estimation Method 2, we survey literature to determine appropriate bounds on all of the $SIPR$ model equations.  We assume that the total number of infected individuals at $t=0$ is 4, as in Ching, et al.~(\cite{Ching2003}). We conservatively assume $0<\beta < 0.01$.  The bounds and constraints for each parameter as well as the estimated parameter values are given in Table~\ref{ParameterTableSARS}.  The resulting number of confirmed cases of SARS are presented in Table~\ref{AmoyGardensTable}.

\subsection{SARS epidemic - large scale}
\label{SARSHK}
We use the $SIPR$ model to study the spread of SARS in the entire Hong Kong Special Administrative Region.  We use the cumulative number of confirmed cases of SARS in Hong Kong as reported by the World Health Organization\footnote{The WHO numbers were accessed on September 29, 2010, on-line at http://www.who.int/csr/sars/country/en/index.html.} (WHO).  The first case of SARS in Hong Kong appeared on February 21, 2003.  The WHO began daily reporting of SARS cases on March 17, 2003, and June 12, 2003 shows the last additional confirmed case to a total of 1755 confirmed cases. In the parameter estimations, we use all of this data, and we summarize the data in Table~\ref{SARSHKTable}.

To begin, we identify three distinct time periods of the SARS epidemic in Hong Kong.  The first time period is February 21 -- March 21, 2003, the second is March 22 -- March 28, 2003, and the third is March 29 -- June 12, 2003.  The first reported case of SARS in Hong Kong occurred on February 21, 2003, when a SARS infected doctor arrived from China.  From this date until March 21, 2003, very little was known regarding SARS and few precautions were taken by individuals to prevent the spread of infection.  On March 21, Hong Kong officials decided to begin daily reports to the public regarding the threat of SARS infection.  As a result of these warnings people began to protect themselves from infection and the dynamics of the disease transmission changed.  Due to the continued spread of SARS and the increasing awareness of the seriousness of infection, on March 29, all Hong Kong schools were suspended, again changing the disease dynamics.  Classes resumed for all students by May 19 and the last reported case of SARS in Hong Kong occurred on June 12, 2003\footnote{A chronology of the SARS epidemic in Hong Kong can be found in Appendix III of the final report by the SARS Expert Committee of the Hong Kong Special Administrative Region, accessed October 15, 2010, on-line at http://www.sars-expertcom.gov.hk/}.

We examine the first and third periods only, using the $SIPR$ model.  The first time period corresponds to quick spreading of the disease and we show this region has an $R_0$ value above 1.  The third period corresponds to the end of the disease and correspondingly has an $R_0$ value below 1.

In 2003, the Hong Kong area had a total population of 6.803 million people.  The number of susceptible individuals in Hong Kong during the SARS epidemic is not the entire population; the number of susceptible individuals must be approximated.  Katriel and Stone~(\cite{Katriel2010}) give a formula to estimate the percent of the population that is susceptible during an epidemic.  The number who are susceptible can be computed using the $R_0$ value and the percent of the population who became infected.  Riley, et al.~(\cite{Riley2003}) determine the $R_0$ value specifically for the SARS outbreak in Hong Kong is in the region 2.2 -- 3.7.  Using this, we compute that the number of susceptible individuals falls in the range 1.8 million -- 3.1 million, where the smaller population correspond to larger $R_0$ values.  We assume a large $R_0$ value, and, therefore we use the lower value, $N=1.8$ million.

For parameter estimation Method 1, we use the parameters determined by fitting the data sets to the classical $SIR$ model.  The fixed and estimated parameters of the $SIPR$ model are given in Tables~\ref{ParameterTableSARSHKone} and~\ref{ParameterTableSARSHKthree}, corresponding to time period one and three, respectively.
The resulting number of confirmed cases of SARS are presented in Table~\ref{SARSHKTable}.

\begin{table*}
\begin{center}
\begin{tabular}{|c|cccc|ccccccc|} \hline
Day & Feb 21 & Mar 17 & 19 & 21 & 29 & Apr 12& 26 & May 10 &24 & Jun 6 & 20 \\ \hline
Confirmed &1 &95  &150  & 203 & 470 & 1108& 1527& 1674& 1724& 1750&1755 \\
SIR  & 1&100.9 &143.8  &204.9  &470.0  &1081.5 &1355.8& 1478.9 & 1534.0 & 1557.6 &1569.3 \\
Method  1  & 1& 100.3& 142.9&203.6 & 470.0 &776.1 &983.6&1124.8&1221.0&1282.6&1328.4\\
Method  2 &1 &103.8 &144.9 &202.2 &470.0  &1156.5 & 1486.1 & 1644.4 & 1720.5 & 1755.1&1773.6 \\\hline
\end{tabular}
\end{center}
\caption{Cumulative confirmed cases of SARS in Hong Kong.   Predicted cases based on parameter estimation for the classical $SIR$ model, and the $SIPR$ model using Methods~1 and~2.}
\label{SARSHKTable}
\end{table*}

\begin{table*}
\begin{center}
\begin{tabular}{|cc|cc|cc|} \hline
Parameter &Dimension & Method 1 &  Fixed  or Estimated & Method 2 & Bound or Constraint\\ \hline
$S_0$ &people & 1,799,999 & Fixed & 1,799,999 &  $S_0$=N-1 \\
$I_0$ &people & 1.0& Estimated & 1.0  &  $I_0 + P_0 = 1$\\
$P_0$ &people & 0.0 & Fixed by $I_0$ &  0.0  & $I_0 + P_0 = 1$\\
$\beta$ &days$^{-1} \times$ people$^{-1}$& 0.2586/N  & Fixed & 0.3738/N  &$0<\beta < 1/N$ \\
$b$ & & 0.6489  &Estimated & 0.8000   &$0.8<b<1$\\
$\nu_1$ &days$^{-1}$& 0.0836  & Estimated & 0.2500 & $\nu_1= .25 $\\
$\nu_2$ & days$^{-1}$& 0.0794 & Fixed by $b$, $\nu_1$, and $R_0$ &  0.1000 &$\nu_2 = .1$\\
$R_0$ & & 3.1511   &Fixed&  1.9438  & $1.5 < R_0 < 4$\\ \hline
\end{tabular}
\end{center}
\caption{Estimated parameters for the $SIPR$ model and the SARS outbreak for February 21 - March 21, 2003, in Hong Kong, using Methods 1 and 2.  }
\label{ParameterTableSARSHKone}
\end{table*}

\begin{table*}
\begin{center}
\begin{tabular}{|cc|cc|cc|} \hline
Parameter &Dimension & Method 1 &  Fixed  or Estimated & Method 2 & Bound or Constraint\\ \hline
$S_0$ &people & 1,799,530 & Fixed & 1,799,530 &  $S_0$=N-470 \\
$I_0$ &people & 277.2 & Estimated & 311.6  &  $I_0 + P_0 = 470$\\
$P_0$ &people & 192.8 & Fixed by $I_0$ &  158.4  & $I_0 + P_0 = 470$\\
$\beta$ &days$^{-1} \times$ people$^{-1}$&  0.1351/N & Fixed & 0.1473/N  &$0<\beta < 1/N$ \\
$b$ & & 0.5237  &Estimated & 0.8910   &$0.8<b<1$\\
$\nu_1$ &days$^{-1}$& 25.6988  & Estimated & 0.2500 & $\nu_1= .25 $\\
$\nu_2$ & days$^{-1}$& 0.0920 & Fixed by $b$, $\nu_1$, and $R_0$ &  0.1000 &$\nu_2 = .1$\\
$R_0$ & &  0.7025  &Fixed &  0.6851  & $0 < R_0 < 1$\\ \hline
\end{tabular}
\end{center}
\caption{Estimated parameters for the $SIPR$ model and the SARS outbreak for March 29 - June 20, 2003 in Hong Kong, using Methods 1 and 2.}
\label{ParameterTableSARSHKthree}
\end{table*}

For parameter estimation Method 2, we survey literature to determine appropriate bounds on all of the $SIPR$ model equations.  We assume that the total number of infected individuals at $t=0$ is 1 for time period one, corresponding to the index case, and 470 for time period three\footnote{Some of these 470 confirmed cases would have already been removed from the disease transmission dynamics via recovery, death, or removal to isolation.}, corresponding to the number of confirmed cases on March 29, 2003. The bounds and constraints for each parameter as well as the estimated parameter values are given in Tables~\ref{ParameterTableSARSHKone} and~\ref{ParameterTableSARSHKthree}.    The resulting number of confirmed cases of SARS are presented in Table~\ref{SARSHKTable}.

\subsection{Summary Results}
We summarize the infections for each of the data sets and their fitted parameters in Table~\ref{ResultsTable}.  Specifically, the table contains the total number of individuals who became infected, the total number who became infected through contact with a regular infected ($I$) and super-spreading individuals ($P$), the total number of regular infected ($I$) and super-spreading individuals ($P$), and the individual reproduction numbers for the $I$ and $P$ classes (as $SIR$ models).

\begin{table*}
\begin{center}
\begin{tabular}{|l|ccc|ccc|ccc|}\hline
 & Amoy&Gardens&&HK & Period 1&&  HK & Period 3&\\
 & SIR & M 1 & M 2 & SIR & M 1 & M 2 &SIR&M 1& M 2\\ \hline
Total Infected &477.3&479.1 & 765.8& 1,709,770& 1,709,769&1,405,410&1578.9&1426.1&1791.5\\
Contact with $I$ &473.3&426.1 &742.1  &1,709,769 &1,097,524&912,542&1108.9&4.1&876.6\\
Contact with $P$ &&49.0 &19.7 & & 612,244&492,867&&952.0&444.9\\
Total $I$ &477.3&443.0 &761.4 &  1,709,770 &1,109,516&1,124,365&1578.9&777.9&1489.2\\
Total $P$ &&36.1 &4.4 &&  600,253&281,045&&648.2&302.3\\
$R_0$ for $I$ &1.5154&1.4779 & 3.4735& 3.1511& 3.0944 &1.4953&0.7025&0.0053&0.5889\\
$R_0$ for $P$ &&2.0255 &8.6838 & & 3.2560 &3.7382&&1.4691&1.4721\\ \hline
\end{tabular}
\caption{Summary of the outbreaks for each of the data sets and their fitted parameters: the total number of individuals who became infected; the total number who became infected through contact with a regular infected ($I$) and super-spreading individuals ($P$); the total number of regular infected ($I$) and super-spreading individuals ($P$); and the individual reproduction numbers for the $I$ and $P$ classes (as $SIR$ models). }
\label{ResultsTable}
\end{center}
\end{table*}

\section{Discussion}
\label{discussion}
We have presented a modification of the classical $SIR$ disease model that captures the effect of super-spreading individuals on an infectious disease epidemic.  Using an idea from the progression of the SARS outbreak, we distinguish the regular infected individuals from the super-spreading individuals by how long they remain outside of isolation; the super-spreading individuals spend longer outside of isolation than most infected individuals.  The model was fit to data from the SARS epidemic, using two different parameter estimation methods.

Parameter estimation Method 1, used parameters estimated for the classical $SIR$ model.  The resulting parameters show only a slight super-spreading behavior in all cases studied.  For Amoy Gardens and the short-term outbreak in Hong Kong, the $I$ and $P$ classes are not distinguished by how long each stays outside of isolation.  They are also not distinguished by their individual reproduction numbers.  In each case, the two recovery rates are similar to eachother, and similar to the recovery rate of the $SIR$ model estimated parameters.  It is not surprising, therefore, that the final number of infected individuals, given by the final number of individuals in the removed class, in the two models are within just a few individuals.

The fitted parameters for the long-term outbreak in Hong Kong using parameter estimation Method 1 indicate that regular infected individuals move so quickly into isolation that we can disregard their influence on the disease spread.  As evidence, over the entire course of the disease, they collectively only infect a total of 4.4 individuals.  Due to the incredibly short time spent in the infected class, the individual reproductive number in this case is very low.

Though the outbreak in the entire Hong Kong Special Administrative Region was split into three time periods corresponding to three different disease dynamics, the fit parameters do not match the third time period data set as well as one might hope.
This is noticed for the classical $SIR$ model, and, since the $SIR$ parameters are used for estimation Method 1, we also see this in the fit parameters for Method 1.
One possible fix for this problem is to split the third time period into other periods in which the outbreak has common dynamics.  

Parameter estimation Method 2, used research to set a priori bounds on the parameters of the $SIPR$ model.  In every case, the resulting parameters show super-spreading behavior, for example, the two infected classes are distinguished by their individual reproduction rates.  All three cases show that, on average, each regular infected individual infects less than 1 other individual, while each super-spreading individual infects more than 1 other individual.  

Considering the specific case of the outbreak at the Amoy Gardens apartment complex, Method 2 predicts that, without any other intervention, almost all of the residents of Block~E, will become infected.  In this case, both the regular infected and super-spreading individuals have basic reproduction numbers larger than one. 
It is clear from the dire outcome predicted for the residents of the Amoy Gardens, Block~E, super-spreading individuals must be brought into isolation as quickly as possible.

For the long-term outbreak in Hong Kong (in the third time period), the overall individual reproductive number is less than 1, which matches the notion that individuals were taking precautions to protect themselves, and the disease spread was slowing down.  Using the $SIPR$ model, we see that the regular infected individuals do have an individual reproductive number less than 1, however, the super-spreading individuals have a reproductive number larger than 1.  Again, we see that it is imperative that the super-spreading individuals be brought into isolation as quickly as possible.

Finally, it is evident from the short-term outbreak in Hong Kong (in the first time period), that without any control measures the spread of SARS in Hong Kong would have been extreme.  The model predicts that more than 1.7 million residents would have been infected over the course of the disease outbreak.  Correspondingly, they would have experienced a large disease related mortality.  The death rate for SARS is estimated to be around 10\%, and so, Hong Kong would have lost an estimated 170,000 residents.  Noting that the final death toll in Hong Kong was (though tragic) only 299 people, the control measures put in place by the government of Hong Kong saved thousands of lives.

The $SIPR$ model is versatile; it can be used to examine an outbreak of any disease known to have super-spreading individuals, measles for example (see~(\cite{Paunio1998})).  On the other hand, the model was built using the idea that super-spreading individuals stay out of isolation longer than regular infected individuals, as during the SARS epidemic.  (There are documented cases of SARS infected individuals violating strict isolation mandates\footnote{See for example ``SARS Epidemic Worsens in Taiwan", by Shu Shin Luh, in the \textit{Washington Post}, May 15, 2003.}.)  For diseases where the super-spreading behavior is a result of differing transmission rates, one should use the Kemper model~(\cite{Kemper1980}).

Super-spreading individuals for infectious diseases pose a serious threat to public health.  The $SIPR$ model clearly demonstrates that infectious individuals must be removed from interactions with the susceptible individuals as quickly as possible to decrease the seriousness of an infectious disease epidemic.


\end{document}